\documentclass[12pt,onecolumn,letterpaper]{article}
\usepackage[left=0.8in,right=0.8in,top=0.8in,bottom=0.8in]{geometry}

\usepackage{times}
\usepackage{epsfig}
\usepackage{graphicx}
\usepackage{amsmath,amssymb,bm}
\usepackage{multirow}

\usepackage{booktabs}
\usepackage{blindtext}
\usepackage{comment}
\usepackage{subcaption}
\usepackage{xspace}
\usepackage{colortbl}

\usepackage{url}

\usepackage{breakurl}
\usepackage[breaklinks]{hyperref}

\usepackage[capitalize]{cleveref}
\crefname{section}{Sec.}{Secs.}
\Crefname{section}{Section}{Sections}
\Crefname{table}{Table}{Tables}
\crefname{table}{Tab.}{Tabs.}

\providecommand{\keywords}[1]
{
  \small	
  \textit{\textit{Keywords ---}} #1
}

\begin{document}

\title{Regulating Multifunctionality}

\author{Cary Coglianese$^{1}$ and Colton R. Crum$^{2}$ \\
$^{1}$ Edward B. Shils Professor of Law and Professor of Political Science\\
Director, Penn Program on Regulation, University of Pennsylvania\\
$^{2}$ Ph.D. Candidate in Computer Science \& Engineering, University of Notre Dame\\
{\tt\small {$^{1}$cary\_coglianese@law.upenn.edu,$^{2}$ccrum@nd.edu}}\\
{\small } \\
{\small Forthcoming in Philipp Hacker, Andreas Engel, Sarah Hammer and Brent Mittelstadt (eds)}\\
{\small The Oxford Handbook on the Foundations and Regulation of Generative AI}\\
{\small (Oxford University Press)}\\
    }
\date{}

\maketitle

\begin{abstract}
Foundation models and generative artificial intelligence (AI) exacerbate a core regulatory challenge associated with AI: its heterogeneity. By their very nature, foundation models and generative AI can perform multiple functions for their users, thus presenting a vast array of different risks. This multifunctionality means that prescriptive, one-size-fits-all regulation will not be a viable option. Even performance standards and ex post liability—regulatory approaches that usually afford flexibility—are unlikely to be strong candidates for responding to multifunctional AI’s risks, given challenges in monitoring and enforcement. Regulators will do well instead to promote proactive risk management on the part of developers and users by using management-based regulation, an approach that has proven effective in other contexts of heterogeneity. Regulators will also need to maintain ongoing vigilance and agility. More than in other contexts, regulators of multifunctional AI will need sufficient resources, top human talent and leadership, and organizational cultures committed to regulatory excellence.\\

\end{abstract}

\thispagestyle{empty}
\clearpage

\addtocounter{page}{-1}
\section{Regulating Multifunctionality}

Many types of tools and technologies perform multiple functions. Consider one of humanity’s most primal of tools: the knife \cite{heinzelin1999environment}. The knife is not a singular tool; rather, it comes in many different varieties that serve many functions, each of which can generate value for society. Knives are used in the kitchen to prepare delicious meals, and then they are used by diners to consume those same meals. Knives carve objects, cut rope, and open packages. They clear paths through forests and jungles, and they help in harvesting seasonal crops. Knives can be used, of course, to injure or kill people. But in the hands of surgeons, knives are routinely used to save lives. And even though knives take many different forms and are often designed for many different purposes—think of, for example, the many types and sizes of surgical scalpels, woodcarver’s chisels, and kitchen implements, among others—knives designed for one purpose also can be adapted for different uses, as anyone who has used a dinner knife to open a postal letter can attest. Many knives, though, are deliberately intended to serve multiple functions, as is the case with a simple pocketknife or, even more emblematically, the classic Swiss army knife, some models of which boast a combination of more than 30 different tools in one.

The proliferation of functions performed by different knives has led over the years to different forms and sources of rules governing their manufacture, sale, and deployment. Many governments place restrictions on the carrying of knives in public when they exceed specified sizes or take a particular form, such as switchblades or butterfly knives. Around the world, airport regulations prohibit the carrying of knives of virtually any kind or size onto the passenger cabins of commercial aircraft. Other rules related to the use of knives take the form of professional licensure standards—such as for surgeons—or private rules of management practice—such as hospital procedures that demand strict accounting of all knives and other surgical tools so that none are accidentally left inside patients. In other spheres, private management standards emphasize training and protocols. Cooking schools provide extensive training in the handling of knives before they award certifications to their students. Youth recreational programs and camps place age restrictions and require adherence to strict safety protocols for when and how camping or craft knives are used. And, of course, overarching all of these other forms of governance, when knives are used improperly or intentionally in ways that harm other people, governmental rules that impose criminal or civil liability come into play.

Digital tools are not entirely unlike their physical counterparts. They, too, call for a variety of different regulatory interventions based on their uses as well as overarching risks. As with knives, AI tools come in many different varieties and serve many different functions that provide much positive value to individuals and society. Some AI tools—such as foundation models—are deliberately multifunctional. They are the digital equivalents of the Swiss army knife. Just as it is hard to imagine every possible use that different individuals might make of actual Swiss army knives, it is hard—even impossible—to specify in advance every single use that can be made of foundation models. Of course, foundation models take multifunctionality to a new level altogether compared with physical tools like the Swiss army knife. Given their extensive range of possible uses, foundation models greatly exacerbate the core regulatory challenge that AI more generally presents: its heterogeneity \cite{coglianese2023regulating}. As AI tools become ubiquitous throughout modern society, their varied nature and highly diverse applications—a heterogeneity that is rapidly expanding and changing—make it increasingly difficult to categorize, define, and subject all the functions that need to be governed.

The result is one of the most daunting governance challenges of our time. How can digital tools that put multifunctionality on steroids ever be subjected to meaningful public oversight and regulatory control so society can be sure that they are used for good, while only posing acceptable, well-managed risks to members of the public? This is the core governance question confronting society in an era of foundation models and generative AI. Regulating multifunctional AI may seem like an impossible task—and it is, unquestionably, a daunting one. But just as the qualities of AI bear an affinity with those of multifunctional physical tools,\footnote{Our analysis in this chapter also applies if one conceives of AI less as a ``tool'' whose \emph{design} needs to be regulated and more as an autonomous ``agent'' whose behavior needs to be regulated. The \emph{behavior} of foundation models can vary widely and fulfill different functions, just as can any \emph{tool}. The analogy, if one sees AI as an agent, is to an expert who performs highly varied tasks. The same hyper-heterogeneity still applies because the ``expertise'' of these models, and the expert tasks they can perform, can take many different forms, as discussed in this chapter.}  the governance solutions available for dealing with the risks of multifunctionality in other contexts also bear similarities to what is needed in the digital realm. Among other things, the governance of AI—even given the extreme multifunctionality of foundation models and new generative AI tools—needs to be multifaceted. It cannot take a single form nor emanate from a single law or lawmaker. In addition, a flexible regulatory structure will be needed to regulate AI’s diverse and ever-changing behavior and risks. 
\section{AI’s Multifunctionality}

The multifunctional nature of AI derives from the demand to use it for highly varied purposes. This ``use heterogeneity'' also drives AI’s ``design heterogeneity''—something which is true for physical tools, too. Knives take different shapes and forms (design heterogeneity) when they are intended for different uses. A surgeon’s scalpel, for example, has a different shape and size than a hunter’s sheath knife. The design of an AI tool also can vary depending on the type of use. But, also as with physical tools, the same basic AI tool can perform multiple functions. Just as it is possible to use a standard kitchen knife to open the wrapping of a birthday present, the same type of AI tool or model—especially a foundational model—can be deployed for different uses. In grappling with AI governance, it is important to recognize the extreme heterogeneity that accompanies AI—in terms of differing designs and, more importantly, across differing uses. It is also important to keep in mind the distinction between tools designed for single or multiple purposes because the intended range of foreseeable uses of a tool can provide regulators the awareness needed to address the tool’s associated risks.

The uses for which AI can be deployed are virtually limitless. It is this use heterogeneity—especially with the multifunctional nature of foundation models and generative AI tools—that presents the most substantial challenge when it comes to regulating AI \cite{coglianese2023regulating, coglianese2016regulating}. Use heterogeneity arises from (a) the existence of a vast array of different AI tools being designed for specific but varied purposes in mind, and (b) AI tools that are themselves designed for multiple purposes, perhaps only some of which can be fully anticipated.

Use heterogeneity arising from (a)—the existence of varied tools designed with different specific purposes in mind—describes many physical and digital tools. Like many knives, for example, traditional AI tools are often designed for a specific purpose, each intended to satisfy a single-function use case. That is not to say that these traditional AI tools oriented toward single uses are feeble or of limited importance. On the contrary, they can prove quite powerful and important. AI tools expressly designed and trained to classify cancerous tumors would fall into the category of a single-function use case, and yet they could hardly be more important. This is not unlike a surgical scalpel that a physician uses to excise a patient’s brain tumor: a vital, albeit single-purpose tool. For these single-function tools (both in the physical and digital realms), it is often possible for regulators or other standard-setters to develop detailed, targeted descriptions of the well-designed tool, such as an AI model’s architectural components or the scalpel’s type of handle, or specifications about how or by whom the tool is used to ensure that it is deployed optimally to deliver benefits and minimize risks.

By contrast, use heterogeneity arising from (b)—that is, from the existence of versatile technologies designed for many different purposes and uses—presents different regulatory challenges. This type of use heterogeneity describes the source of the challenges associated with regulating the foundation models and generative AI that constitute the focus of this volume. Although the bulk of AI tools that have been developed and deployed until recently have consisted of algorithms designed and trained to fulfill singular tasks, it has also long been the hope that AI would be able to address more generic, real-world problems by performing multiple functions and tasks, much like humans do. Progress has accelerated to a point where, today, digital tools can handle a variety of tasks at or even exceeding human levels, even though AI is not (yet) fully generalized to the degree of the broad range of humans’ own inherent multifunctional abilities. Today’s AI models, bolstered by an intense deployment of computational resources and a prodigious effort to excavate mammoth amounts of data and curate them for model training, have improved not only the performance of single-function tasks but can now advance a sweeping range of multifunctional tasks downstream, far more than their designers expected. 

A paradigm shift has occurred: AI has become akin to the Swiss army knife, made intentionally to handle a virtually uncountable range of uses with passing, if not even remarkable, proficiency. As a result, a necessary distinction today can be made between single-function AI and multifunctional AI, with the latter aptly named ``foundational models.'' Foundation models are defined as any machine learning or deep learning model that is broadly trained to handle a wide range of tasks downstream, including many unknown to the developer beforehand \cite{bommasani2021opportunities,team2023gemini,achiam2023gpt,touvron2023llama}. Like the Swiss army knife, foundation models are developed to be generic and sufficiently general to serve many (if not almost all) desired purposes of the end users. The term ``foundation model'' may also be colloquially used interchangeably with adjacent terms, such as ``frontier,'' ``general-purpose,'' ``generative AI,'' or ``large generative AI models'' \cite{hacker2023regulating}. Many of the large language models (LLMs), such as ChatGPT, fall into this category. Regardless of the nuances in terminology, these forms of AI all fall under an umbrella of what we consider here to be ``multifunctional AI,'' with design choices and regulatory implications that are distinct from ``single-function AI.''

With the number of AI applications expanding rapidly—both the single-function and multifunctional varieties—the number and novelty of AI uses have skyrocketed. AI is being deployed by private firms to perform a plethora of industrial functions (e.g., robotic manufacturing, inventory control, and more) and to power a range of consumer-oriented products (e.g., antilock brakes and self-driving functions in automobiles, social media platforms, and more). Medical professionals use AI for tasks as varied as reading radiographs, customizing medical treatments, and summarizing clinical notes. Legal professionals are finding AI useful for document review and writing tasks. Computer software engineers are relying on AI tools to augment and streamline their programming. In the public sector, AI tools are helping law enforcement and regulatory agencies identify possible sources of legal violations \cite{coglianese2016regulating,coglianese2020ai} as well as helping municipalities decide where to target limited social services resources \cite{engstrom2020government}. One study has identified over 1,300 nonmilitary use cases by the U.S. federal government as of 2023 \cite{coglianese2025fu}. This number will only increase. Moreover, the use heterogeneity of so-called traditional AI (that is, different models performing single functions) pales in comparison with the uses to which individuals are putting LLMs and generative AI tools: composing emails, researching trivia, crafting recipes, producing entertainment, developing digital friendships, and so much more. With ``hundreds of millions of weekly users,'' large language models are regularly being deployed to serve what one study identified as at least a hundred different categories of uses \cite{zao2024genAI}. As the authors of that same study noted, ``[t]he use of this technology is as wide-ranging as the problems we encounter in our lives'' \cite{zao2024genAI}.

The upshot is that the uses of multifunctional AI tools are essentially limited only by the imagination of their end-users. Most generative AI models rely upon a user-requested prompt \cite{team2023gemini,achiam2023gpt,touvron2023llama}. Slight changes in the prompt will elicit different model responses, and the user can probe and shape the AI’s output through so-called prompt engineering. Writing effective generative AI prompts is now a marketable skill: companies are requesting prompt engineers with salaries as high as \$$335,000$ \cite{quilty2024}. Despite not having any internal access to the model’s internal mechanisms, users can alter the behavior of generative models simply by carefully worded prompts. Although AI companies have implemented so-called guardrails to prevent their models from answering certain questions (illegal, offensive, or self-harm-related questions), their protective methods can sometimes be circumnavigated with a few cleverly worded prompts or so-called jailbreaking techniques \cite{shen2024anything}.

The incredibly varied uses to which AI can be deployed can be both productive and risky, socially beneficial and downright pernicious. Specifying with precision all the benefits and costs of foundational models and generative AI would be an impossible task. Nevertheless, when these models form the backbone of task-related tools—such as when LLMs are used to power legal-drafting or medical note-taking programs—the experienced user or domain expert can be more discerning of the likely range of uses and the tools’ performance. Developers of foundation models can also observe a broad range of foreseeable uses, even if they are numerous. They may even be able to use AI tools to forecast or analyze the range of uses of their general AI models. That said, having the foresight to anticipate every potential use is surely impossible, just as it is with physical tools. Nevertheless, for regulators to establish effective modes of managing AI’s risks, they must wrestle with a third type of heterogeneity that is endemic to AI tools. In addition to \emph{design} heterogeneity and \emph{use} heterogeneity, AI tools exhibit \emph{problem} heterogeneity. By ``problem heterogeneity,'' we simply mean the wide array of distinct risks that follow from use heterogeneity—whether by varied AI designs intended for single functions (traditional AI) or by multifunctional or foundational AI tools (generative AI).
\section{The Multifaceted Governance of Multifunctionality}

The heterogeneity of problems posed by AI creates a core challenge for AI governance. This is true for both varied single-function AI tools as well as for multifunctional AI tools. The problems presented by a specialized AI tool that misreads MRIs used to identify cancer, for example, will obviously be quite different from those presented by a specialized AI tool assisting in the automatic braking of an autonomous vehicle. Likewise, the problems created by an LLM tool ``hallucinating'' about legal cases will be different from these same foundational LLM tools put to use in social media that contribute to teenagers engaging in self-harm. The question thus becomes how society should approach regulation when both \emph{uses} and \emph{problems} can be so varied. In other words, how does society regulate the AI equivalent of the Swiss army knife?

It should be recognized initially that, among the array of problems arising from AI, many can be grouped into categories that reflect classic market failure problems that have historically necessitated regulatory intervention: market concentration, information asymmetries, and externalities. When AI is used to power automated pricing mechanisms by online retailers, for example, it may lead to anti-competitive practices, distorting the market and harming consumer welfare \cite{justice2024}. When embedded in health care products and consumer goods, AI tools can create information asymmetries, as consumers will not be as aware as manufacturers of any safety risks. And the risks posed by AI-enabled self-driving cars illustrate how AI tools can create the kind of externalities that typically justify some regulatory oversight.

In addition, AI tools can present concerns beyond those considered to be classic market failures. When models are trained on biased data that reflect societal prejudices, they can perpetuate and even amplify existing inequalities. This risk is particularly acute in employment contexts, where algorithms used in hiring processes may unintentionally reinforce discriminatory practices. AI tools also pose a range of other problems or potential problems, such as those arising from the amount of energy they demand, their potential for unfairly appropriating copyrighted material, and their risks of making major disruptions to various segments of the labor market. Privacy concerns also exist, as the vast datasets leveraged by neural networks and other AI models often contain sensitive personal information. The potential for these models to infer private attributes not explicitly included in the data can also raise additional concerns about privacy and the protection of individual autonomy. Furthermore, the threat exists that AI will help enable a surveillance state, especially when technologies such as facial recognition are widely deployed.

Added to this litany of concerns are a host of distinctive public policy questions and challenges raised by large language models and other generative AI tools. Given how similar the writing produced by generative language models compares to that of humans, end-users often forget they are not always accurate or operating with true empathy. Generative AI models can lack the ability to understand basic common sense and logic. They can be prone to hallucinations or false conceptions of reality, and they may make up fake things, such as historical events, court cases, or book titles. Such inaccurate information can then easily spread on social media, contributing to disinformation and misinformation campaigns. In March 2022, for example, a widely circulated AI-generated (and thus fake) video of President Volodymyr Zelenskyy of Ukraine telling Ukrainians to lay down their weapons and surrender during the Ukraine War was shared on social media, causing public alarm \cite{simonite2022}. Generative AI also raises concerns about recreating some of its training data, which could ``leak'' personal data in its output \cite{tinsley2021face}. In addition, foundation models have been prone to use hate speech, encourage racist behavior, hold racial biases, threaten to harm users, and incite malicious and inappropriate behaviors \cite{perrigo2023}. Although some of these problems may have relatively straightforward fixes, retraining algorithms can sometimes take months and cost a model developer large sums of money. Data scientists are undertaking active research into model ``unlearning'' for this very reason \cite{neurips-2023-machine-unlearning}. 

In other contexts, generative AI raises still other concerns. For young students, generative AI has the potential to disrupt conventional modes of learning and assessment. The AI models that support social media platforms may unintentionally (or, at times, perhaps intentionally) fuel political polarization, harm youth, and foster addictive behaviors. The creative industries face their own challenges with AI, as generative models capable of generating new art from existing works blur the lines of intellectual property and ownership, thereby altering incentives for human creativity. Worries about replacing human creators and other workers have become palpable throughout a variety of sectors of the economy. In addition, malicious uses of AI tools are ever-present, such as their use in the orchestration of sophisticated cyberattacks or the creation of deepfakes that are then used to manipulate voters and swing elections.

Taken together, the many problems that AI tools can create may well be nearly as varied as the tools and their uses themselves. Given that generative AI and foundational models have the ability to be used in innumerable ways, and given that they possess the capacity to generate any form of output in response to prompts that were not included in their training, there comes with each of these models the potential for novel risks and harms. Importantly, just as with multifunctional physical tools, such as knives, the problems associated with AI are not necessarily inherent in, and thus identifiable only from, the technology itself. Instead, both the productive and harmful effects of multifunctional AI depend on the uses to which this technology is deployed. By analogy, the same kitchen knife that can be used to dice a tomato in preparation for a Friday night dinner with a spouse could also be used to open a birthday package sent to the same spouse or to protect that spouse from a home intruder. Likewise, any of these acts or uses could be performed using a consumer-grade steak knife purchased at a local grocery store, an industrial-grade paring knife designed for the finest of chefs, or even a makeshift knife forged from a broken wine glass. 

The context around AI models and their uses matters, too—again, just as with physical tools. After all, dicing a tomato is trivially safe for an adult, but it could be extremely dangerous for a child. Dicing a tomato is also inconsequential for adults who are cutting in well-lit kitchens, but not for those cutting in the dark. Environmental conditions matter. Foundational and generative models exhibit similar properties and contextual challenges, each dependent on the varied circumstances surrounding their varied uses. Accomplished lawyers, for example, should be able to detect when an LLM is hallucinating and making up legal cases, but prison inmates representing themselves \emph{pro se} will be unlikely to do so.

One thing is clear: multifunctional technology (physical or digital) demands multifaceted regulation. No single, magic-bullet solution exists. Likewise, no single sweeping piece of legislation or regulation, nor even a single regulatory body, will be sufficient. Furthermore, the human capital required to govern AI responsibly will need to be expansive and even heterogeneous. No single form of expertise will be adequate, given the sheer degree of AI’s multifunctionality. Responsibly regulating AI will require an amalgamation of effort from many organizational bodies taking on an ``all-hands-on-deck'' approach. It will necessitate reliance on the different experiences and forms of expertise residing in a variety of seasoned regulatory bodies and standard-setting organizations as well as in the private-sector firms that build these powerful tools and even among these tools’ public end-users, who can provide feedback based on their experiences. The pairing of diverse human capital with flexible regulatory approaches defines what must ultimately be a \emph{multifaceted governance of multifunctionality}.
\section{The Ill Fit of Prescriptive Regulation}

Just as multifunctional physical tools such as knives can be used in both beneficial and harmful ways, so can AI tools. When any tool or technology delivers benefits but comes with side effects or risks, it becomes a prima facie subject for regulation. Regulation seeks to manage risks, not necessarily to prevent them entirely. After all, consider the loss that humanity would incur if all knives and other sharp objects were banned instead of regulated simply because they also posed some residual dangers. In a similar manner, it would surely be undesirable, even if it were possible (which is doubtful), to ban foundation models and generative AI outright, purely because they can pose risks. But just as different risk management rules—public and private—have arisen around the use of physical tools, such as knives, we should expect to see an array of regulatory efforts to manage the risks posed by AI tools. The same basic array of approaches taken in regulating multifunctional physical tools and other risky technologies and behaviors can be applied to regulating foundation models and generative AI tools. 

One common regulatory strategy, though, will prove especially difficult when it comes to regulating the multifunctionality of foundation models: the so-called prescriptive strategy.\footnote{All regulation can be thought of as ``prescriptive'' in some sense, as it inherently prescribes or imposes some kind of obligation on regulated entities or individuals. In this chapter, we rely on a common association of the label ``prescriptive'' with what a National Academies of Sciences, Engineering, and Medicine committee has more clinically and helpfully referred to as ``micro-means'' regulation \cite{national2018designing}.}  Prescriptive regulatory measures—sometimes called micro-means rules \cite{national2018designing,coglianese2024rule}—impose fixed mandates on the design of products or the actions or behavior of regulated entities, including in their development or use of products. With AI, for example, this would mean specifying exactly how algorithms need to be designed and trained—such as what particular algorithmic forms and mathematical functions would be permissible to be used or how different algorithms could be used. This prescriptive approach will prove to be both unrealistic and inapplicable to governing multifunctional AI for several reasons.

For one, as discussed previously, AI’s use heterogeneity makes it unrealistic that regulators would ever have enough information to tell the firms developing AI exactly how their explicit digital tools can and cannot be designed or used. Moreover, regulators would need to foresee all the uses of an AI tool’s client or user base. With foundation models, the sprawling and often surprising uses (including both beneficial and problematic) that the public has found since the release of OpenAI’s ChatGPT reveals the limitation of a prescriptive or micro-means regulatory strategy for multifunctional AI. How could any regulator issue enough rules by commanding specific actions to cover all the possible uses? Furthermore, the \emph{problem heterogeneity} that follows from \emph{use heterogeneity} creates similarly daunting challenges for prescribing all the specific means that should be adopted. If the firms developing an AI tool are often surprised by its competence to handle novel problems, then one can reasonably ask how regulators are going to forecast these tools’ potential problems with enough specificity to enable them to embed into rules specific actions that the developers must take to avoid these risks.

In addition, prescriptive regulation typically works in circumstances where one-size-fits-all solutions can be found. This might sometimes be possible with fixed, single-function technologies, but it clearly does not fit with multifunctional AI. Indeed, regulating foundational AI is far more daunting than simply regulating a standard Swiss army knife, as the same multifunctional AI tool can itself effectively change and adapt after it is initially created. Multifunctional AI is, in essence, a Swiss army knife that not only sharpens its own blade but can lengthen or change the shape of its blade depending on what is necessary for the present use case. This level of environmental adaptation seems to be unlike any technology previously created, further elevating AI’s autonomy such that it behaves more like an animal (or a human being) than a mechanical artifact.\footnote{These subtle shifts in AI continually challenge our concept of intelligence. Yoshua Bengio, a Turing Award laureate who is considered by many observers to be one of the ``Godfathers of AI,'' has expressed that AI that surpasses human intelligence will do so by being better at a ``vast array of tasks'' \cite{bengio2023}. If intelligence (either human or artificial) is surmised by it being a ``versatile task-completion machine'' \cite{vallor2024danger}, then regulating foundation models may look less like regulating products and more like regulating behaviors. That is, as indicated \emph{supra} in note 1, the challenge of AI regulation may be more akin to regulating behaviors of both individuals and organizations.}

Prescriptive regulation of multifunctional AI will also be generally infeasible because the technology itself is varied and dynamic. Like the many different types of knives, there are many different types of AI models—i.e., they exhibit \emph{design} heterogeneity. Neural networks are by far the most prevalent type of AI model, serving as the bedrock of contemporary AI (both single-use and multifunctional). Neural networks comprise a series of interconnected neurons, organized in ``layers'' that process inputs and relay outputs to other neurons. How these neurons are arranged (and in what order) determines different AI architectures and models (e.g., convolutional versus transformer neural networks).\footnote{\emph{Convolutional} neural networks in vision contexts, for example, use pre-trained weights as image ``filters'' that indicate representations of edges, textures, and colors into the model \cite{lecun1989backpropagation}. \emph{Transformer} neural networks, which were popularized by the GPT-series, attempt to learn the ``attention'' between a given neuron and previous neurons \cite{vaswani2017attention}.} Even within the same type of neural network, the neurons themselves can be configured in different ways.\footnote{For example, residual functions or ``blocks'' (known commonly as ResNet) can be added to any neural network layer, and subsequently will change any neural network’s overall architecture \cite{he2016deep}.}

Despite the versatility of neurons and their algorithmic arrangement, neural network models can be trained for virtually any task and domain, although some architecture types favor some tasks over others. Convolutional neural networks have long dominated computer vision tasks \cite{lecun1989backpropagation, krizhevsky2012imagenet}, but they have also been successfully used in a variety of applications, ranging from drug discovery to high-level strategy games such as Go \cite{wallach2015atomnet,silver2017mastering}). Conversely, transformer neural networks are superior in natural language processing (NLP) \cite{vaswani2017attention}, but they have also been used for wider tasks such as protein folding \cite{abramson2024accurate} and, more recently, computer vision \cite{dosovitskiy2020image}. Transformer neural networks were initially designed for language tasks, but their emergent behaviors have led engineers to try them with vision-related tasks, to considerable surprise at their capabilities \cite{vaswani2017attention, dosovitskiy2020image}.

Unlike with knives, making the correct AI selection in type and architecture is only part of the challenge. An AI model must be appropriately taught how to complete a task, creating what might be considered to be training heterogeneity. Algorithmic training is by far the most important step in creating any AI tool. For instance, the training of neural networks involves undertaking a computational regimen by which each neuron adjusts to satisfy or solve a specific task (e.g., to identify a tumor).\footnote{The number of neurons adjusted during training depends on the difficulty of the task and available training samples, ranging from tens of millions (ResNet50, computer vision tasks) to hundreds of billions (especially with the GPT-series LLM) \cite{bommasani2021opportunities,he2016deep}.} Of course, AI models can ``solve'' a desired task but sometimes do so in an undesirable way, such as by latching onto extraneous information correlated with class labels (such as timestamps) or making unfounded correlations that still satisfy the overarching training condition. In an effort to minimize these errors, the tuning of neurons can become so complex that, holding all factors constant, it is impossible to create the exact same model twice, making each trained model completely unique. Nuances in computational hardware or changes to a single neuron can have unforeseeable consequences—again, making infeasible any effort by a regulator to specify any ex ante prescriptive rules.

Regardless of an AI tool’s model architecture or its training configuration, it can only extract inferences from the data used to train it. These models do not have the capacity to learn from the most current data or events because they are trained on data from the past. Because the world, and data about it, can be in continuous evolution and updating, this fluidity only contributes further to AI’s training heterogeneity. Slight changes in or updates to data can sometimes affect a model’s behavior. Changes include those related to collection, file formatting, class imbalances, and data compression techniques. Models can also learn undesirable traits that lurk within the depths of their training data. For example, LLMs might generate harmful, offensive, or racist outputs based on the biases embedded in data accumulated by human systems \cite{bommasani2021opportunities, bolukbasi2016man}. For this same reason, AI models have sometimes been shown to display common biases, including confirmation, selection, group attribution, contextual, and anchoring biases \cite{bommasani2021opportunities, buolamwini2018gender, gallegos2024bias, hullman2022worst}. Engineers need not explicitly instruct the model to learn this information, but some models are able to extract these undesirable traits simply by their existence within the training samples \cite{bolukbasi2016man}.\footnote{Also, collecting accurate, or ``ground truth,'' labels may evolve. Elementary classification labels involved in classifying animals (e.g., dogs and cats) can be rather straightforward, but in some instances even these can change. In medicine, for example, a diagnostic procedure may change the classification of a disease as clinical definitions and technology (e.g., radiology technology such as PET scans) develop over time. As societal definitions change (including cultural terminology), AI models must be given sufficient contemporaneous information to retrain. But they will then effectively become new ``models.''} 

The upshot is that AI is far from a single ``technology'' but instead it comprises a variety of different types of models, architectures, and training processes. This is not to deny that there will be ``best practices''—or at least better means—when it comes to designing AI tools. It is just to say that with this kind of heterogeneity in models, their training, and their underlying data, it is far from realistic to think that regulators will be able to specify exactly how to build (or how not to build) foundation models, or any type of AI. As a practical matter, and generally speaking, prescriptive regulation simply will be infeasible. 
\section{Flexible Regulatory Approaches}

Instead of rigid, prescriptive rules, the future of AI regulation will likely depend on more flexible regulatory strategies. At least four more feasible regulatory strategies exist that could be considered for the governance of multifunctional AI: performance standards, disclosure regulation, ex post liability, and management-based regulation. These are by no means panaceas, as the conditions for their effective use will not always apply. Still, they are more promising in the face of AI’s hyper-heterogeneity than so-called prescriptive rules. Moreover, regulators can rely on a combination of these strategies depending on their overarching goals and context. Importantly, none of these strategies necessarily demand a ``one size fits all'' approach. 

\subsection{Performance Standards}
Performance standards could be conceivably used for known and well-defined problems with specific applications of AI. These standards will thus be more likely to apply to AI tools that are intended to serve a single or limited set of functions, where outcomes can be adequately measured and defined with precision. AI tools that might suitably come under performance standards will typically be those of the more ``traditional AI'' variety. However, even multifunctional AI models, such as LLMs, may be re-purposed to serve a single function—akin to a Swiss Army knife only being used to open mail. Although it may make more sense from the manufacturer’s point of view to change knives, the ease of grabbing the digital equivalent of the Swiss army knife—foundational models—``off the shelf'' has led many users to rely on them for singular functions. For example, a foundational model may be tasked to take patient information and respond with medical advice. Or it may be incorporated into a legal research tool to support specific tasks such as citation checking or contract compliance detection. Performance standards might well be applicable within these narrow and well-defined contexts, even though multifunctional AI serves as the foundation for these specific function tools.

Of course, once the purpose or function of an AI tool changes or increases in complexity, performance standards will be less viable. Consider the success of a foundation model built into a digital system used by a restaurant to take orders and relay them to the kitchen staff. That use may seem benign. But if that same foundation model is then repurposed to answer customer questions related to life-threatening allergens, the consequences could be significant. If the same model is further used to communicate in other ways with customers, it could potentially respond with offensive answers. Performance standards are by necessity designed for identifiable problems. Unless all problems can be specified by the regulator in advance, this approach will be incomplete. Performance standards also necessitate the existence of some way for regulators to test or monitor for compliance, which might be exceedingly difficult to ensure, even when suitable problems can be anticipated and standards defined \cite{coglianese2016limits}. These limitations may, in many if not most cases, make reliance on performance standards as infeasible as reliance on prescriptive or micro-means regulations \cite{coglianese2023regulating}.

Firms, rather than regulators, will generally be in a much better position (assuming they have the right incentives) to understand the ways in which users can use their technologies, the broad range of problems associated with the tools they develop, and how to monitor for the presence or level of those problems. Furthermore, an algorithm’s design might be thought of as a performance ``standard'' itself, as it is through this design that firms can effectively ``embed'' value choices in ways that keep AI tools in check. But just as with prescriptive regulation, regulators will be clearly at a disadvantage in instructing anyone about how algorithms should be designed, which will also necessarily imply limits on their ability to rely on their own externally crafted performance standards. Regulators’ main challenge will likely instead be to ensure that firms use their comparative advantage to establish their own internal ``performance standards'' to protect consumers and the public from undue harm from their tools—an approach we discuss further below in the section of this chapter on management-based regulation.

\subsection{Information Disclosure Regulation}
In the face of the information asymmetries associated with advanced technologies, regulators can require that firms disclose information. Information disclosure regulation may enable regulators to do better by way of monitoring the risks associated with AI technology and potentially provide an incentive for firms to improve their management of those risks \cite{kleindorfer1998informational}. With respect to AI, two main types of information disclosure regulations are plausible. 

The first type might be called existence disclosure regulation. Under this type, consumers and other end-users must be informed that they are interacting with a system that relies on an AI tool—that is, what must be disclosed is that AI exists as part of end-users’ interaction with or use of a larger product or service. After all, unlike with a knife, AI is commonly invisible to the end-user, frequently operating behind the scenes. By demanding the disclosure of the existence of these elusive uses of AI, regulation may advance both rights-based and instrumental values. This may allow consumers and others to be on their guard against errors and report them to the regulated firms or to regulators. Such regulation may require the disclosure of basic information about the AI’s design as well as details about how a consumer could report any undesirable results or concerns.

A second type of disclosure requirement might be called a performance disclosure regulation. This type of disclosure is similar to performance standards. But instead of (or perhaps sometimes in addition to) mandating the attainment of any actual levels of performance to be achieved, performance disclosure regulation simply demands that evidence of performance be disclosed. For example, the performance of AI tools on various tests—both for their benefits and potential harms—could be required to be disclosed through model or system cards \cite{mitchell2019model, gebru2021datasheets, richards2020methodology}. These performance disclosure cards or reports are sometimes compared to the equivalent of nutritional labels required to appear on food products. Such performance disclosures in the AI context might plausibly create incentives for improvement of a tool or a business’s use of it. If firms must disclose their AI tools’ performance on various tasks, consumers and others can better assess risks and make informed choices about the possible harms associated with different AI tools. When the required disclosures occur in a standardized form, and are based on common tests, consumers can better compare different tools, with the mandated disclosures providing a level playing field for market competition. Competition on a level playing field might provide incentives for AI firms to improve their management of AI risks. 

As appealing as performance disclosure regulation might be, it can suffer from many of the same challenges associated with performance standards themselves. This type of disclosure regulation depends on a well-defined understanding of the problems created by AI tools and an ability to specify how to measure and report these problems. Firms may have incentives to game these requirements—either by creating entirely false performance test results or, more benignly but still problematically, finding ways to score well on the tests even while not doing as much to reduce the real problems \cite{coglianese2016limits}. In addition, mandated performance disclosure comes with additional challenges. If it is to serve as a driver of AI safety and risk management, its impact will depend to a considerable degree on consumers and others being able to process the disclosed information. At least as model cards are presently conceived by AI firms and data scientists, these disclosures are often replete with technical details that are difficult for ordinary consumers to understand. If regulators choose to rely on information disclosure as a regulatory strategy, they must think hard about the optimal disclosure technique (e.g., textual descriptions, audible alerts) and the content of the information (e.g., model cards, warning labels) that must be disclosed.

\subsection{Ex Post Liability}
Another flexible regulatory tool for governing AI involves the development of standards for ex post liability once problems with an AI tool arise. Given multifunctional AI’s hyper-heterogeneity, ex post liability must surely be considered as part of an overall multi-faceted approach to AI governance. It is not surprising that there have been many calls to implement ex post liability, whether through AI-specific insurance compensation schemes \cite{lior2021insuring, lior2023innovating} or by reliance on tort law \cite{calo2015robotics,crootof2015war,crootof2017international,karnow2016application,marchant2012coming}. Ex post liability may be advantageous because it does not demand that a regulator know in advance whether a general LLM will cause damages from errors in data collection or cleansing (e.g., biased data), specific architectural components (e.g., the algorithmic design), or training configurations—or from fine-tuning or even hardware problems. Nor does the regulator have to know with precision in advance how to define the precursors to a harmful event as would be needed to develop measurable performance standards.\footnote{Sometimes regulators will adopt binding but general and unmeasurable performance standards in the form of broad principles to which regulated entities must comport \cite{black2008forms, schuett2024principles}. Although such a so-called principles-based approach to regulation is sometimes characterized as ``performance-based,'' the lack of precision and measurability means that this approach, when it creates binding legal obligations, operates in a manner akin to ex post liability rather than as the measurable performance standards discussed earlier in this chapter. For example, AI legislation in Brazil \cite{pacheco2023projeto,zanatta2024rielli} articulates a series of principles about which it has been said that, ``technically, any company directly or indirectly infringing them can be fined'' and where ``direct or indirect non-compliance with the list of principles might also lead to liability infringement'' \cite{jarovsky2024principle}. This is an example of binding principles-based regulation. It is possible, of course, for regulators also to articulate principles as a form of non-binding guidance or soft law. Marchant and Gutierrez \cite{marchant2022soft}, for example, offer a thoughtful discussion of the role that non-binding principles and soft law standards could play in AI governance. Principles-based strategies of this latter type may be important but they are outside the scope of this chapter, which focuses on options for binding legal obligations—that is, for regulations.}  

Although useful, ex post liability may be insufficient as the sole form of multifunctional AI governance. Such liability only applies after harm has been inflicted, when what society often seeks is a more preventative regulatory intervention. Liability also can depend on convincing another entity—such as a court—that a firm acted improperly, which introduces uncertainty that may blunt the ex ante, preventative effects of a liability standard. 

Another challenge rests in defining who the liable party is: the designer and developer of the foundation model; the source of the underlying data on which the model was trained; a firm that adapted or made use of the foundation model in some other tool; or the user who applied the multifunctional tool improperly or to ill effect. If ex post liability hinges on causation, it may be difficult to establish the causal connection between the behavior of the AI’s creator and the plaintiff’s damages \cite{bathaee2017artificial}. Since AI models are largely free to solve a task in whichever way they see fit (attempting to minimize the loss function during training), model behaviors do not always reflect human intuition, which may make it difficult for plaintiffs to prevail under liability standards defined in terms of a ``reasonable person'' and ``foreseeable'' harm \cite{coglianese2021algorithm}. 

Another limitation of strict applications of ex post liability could be that they blunt many potential positive effects of AI, not only the negative ones. The possibility of overregulation is one of the concerns about outright bans on AI use \cite{satariano2023}, and this concern can apply to the application of highly stringent forms of ex post liability as well. Nevertheless, given the rapidly expanding reliance on AI throughout the economy, its attendant harms probably cannot escape all calls for the ex post assignment of responsibility. That said, neither need the public expect regulators simply to wait for problems to develop before taking action. 

\subsection{Management-Based Regulation}
A more realistic preventative strategy for regulating multifunctional AI would embrace its hyper-heterogeneity and apply rules that leverage the private sector’s ability to provide a frontline defense against its risks. Management-based regulation is arguably the only regulatory strategy equipped to handle the heterogeneity challenges posed by foundational and generative AI. This type of regulation is already used widely in other diverse and dynamic risk settings where neither prescriptive nor performance standards are feasible \cite{coglianese2023regulating,coglianese2003management, coglianese2010management}.

Under a management-based approach, a regulated entity is required to develop an internal plan to identify and monitor risks, establish and implement protective procedures to manage those risks, and document changes made to keep addressing those risks over time. A management-based approach applied to multifunctional AI would obligate AI developers to think deeply about and gather information on the broad range of potential uses of their multifunctional tools and subsequently take those uses into account when iterating over their design and development phases. Internal plans would facilitate these reflections and provide a clear path for remedying any potential roadblocks that lay ahead. These plans would need to articulate specific appropriate internal practices and procedures (e.g., validation, auditing, red teaming) to identify and manage risks. Throughout a required iterative, ``plan-do-check-act'' process, AI firms would establish organizational routines through which they relay known issues to their users—and allow users to relay them to the firm—and make necessary adjustments to their models or their risk management practices.

Regulators can subject the required plans and their implementation to routine auditing—by third parties, government overseers, or both. They can enforce the periodic updating of planning protocols and internal risk management practices as new information surfaces about algorithms’ use, training configurations, and ultimate performance. In extreme cases, regulators could undertake emergency audits if made aware of potentially dangerous ``bugs'' or defects that may present themselves in certain model architectures or algorithmic training procedures. Overall, the level of management scrutiny demanded will depend in part upon the multifunctional AI tool’s capabilities (as with rules that vary depending on the length of a knife) and the risks surrounding the tool’s environment (such as with knives’ presence in passengers’ carry-on luggage on commercial airplanes). 

Regulators can require firms to assemble technical information both to facilitate internal management and external auditing. Some of this information could appropriately become part of a public disclosure strategy, as noted above. But management-based regulation recognizes that the technical nature of what firms might need to assemble might prove of less value as part of an information disclosure regulatory strategy than as part of ongoing managerial oversight. In other words, by requiring that managers assemble and systematize information on their AI design and development processes, those managers (and regulators themselves) will be better positioned to track down leads, cross-examine technical issues, and spot global trends that may potentially arise in certain AI architecture, hardware, or data sets.

Management-based regulation can be used in combination with other regulatory strategies at the same time that it also addresses several ongoing limitations of other forms of regulating multifunctional AI. It eliminates the need for regulators to possess detailed knowledge about each AI tool developed or used by every firm. Given the accelerated pace of AI development, it can prevent regulatory decay that is likely to occur with any micro-means rule or performance standard. Regulators would still obviously need to work alongside firms to stay up to date with current market trends, industry changes, and newfound risks. But firms would have flexibility to pursue technological innovation, while regulators would have a basis for monitoring innovations and encouraging firms to address emergent risks.
\section{Regulatory Vigilance and Agility}

Just like a surgeon choosing the appropriate knife, or a firm selecting the appropriate AI chatbot for customer service, regulators must select the ``right'' regulatory tools to fulfill their vital mission. But selecting a particular regulatory strategy—or a combination of strategies—will be only the beginning. Perhaps the most difficult work in regulating multifunctionality will lie in providing ongoing vigilance and agility. Effective AI governance will require constantly adapting, issuing alerts, and prodding action. Regulators need to see themselves as overseers of dynamic AI ecosystems, staying flexible, remaining vigilant, and always seeking ways to improve.

Maintaining regulatory vigilance and ability will demand resources: financial resources, technological tools, and, perhaps somewhat ironically, human capital \cite{metzenbaum2015makes,coglianese2024people}. Human oversight of complex, flexible regulatory systems is essential. Without fervent attention to changes in AI’s uses and problems, new risks can emerge and threaten the public. Regulatory slippage can also arise, especially if private firms’ managerial efforts relax over time in the absence of ongoing attentiveness by government regulators. 

The heterogeneity ushered in by multifunctional AI will prove more taxing on the attention of regulators compared with conventional technologies. It will also necessitate further diversity and inclusion of human expertise within regulatory organizations as well as across other so-called stakeholders. Regulators will benefit from finding optimal ways to collaborate with firms, think tanks, academic centers, standards-setting bodies, advocacy groups, and members of the general public. 

Vigilance will also be necessary from the end-users and greater public, who often find themselves on the receiving end of the negative effects of these technologies. End-users can play an active role in helping to shape these technologies that can impact their lives if they are given a meaningful voice in AI governance. Notable instances of biases, flaws, and limitations of LLMs have been discovered not by the firms that have built or deployed these models, but instead by users \cite{perrigo2023}. Providing users with reporting avenues to channel their input into proper firm-based feedback loops can help guide necessary iterations to improve AI.

Ultimately, a persistent challenge remains in maintaining regulatory excellence in light of the blistering pace of AI development and deployment. Successful regulatory agility in governing multifunctional AI will involve both ongoing attentiveness and an ability to act quickly when needed. 
\section{Conclusion}
As powerful, multifunctional digital tools, foundation models and generative AI tools operate like Swiss army knives. They offer a seemingly limitless number of potential uses and create an array of exciting opportunities when used safely and fairly. But they are not without their risks. Furthermore, they exacerbate the core challenge in regulating AI: its heterogeneity. AI’s heterogeneity begins with its numerous uses and extends to the number of potential problems it can create. It is AI’s heterogeneity of uses and problems that makes it akin to a very expansive, multi-tooled Swiss army knife. Although regulating multifunctionality is a daunting challenge, regulators can meet that challenge by taking a multifaceted approach. Much as physical tools—such as actual knives—fall under different sets of rules depending on their context, AI tools can also come under varying regulatory regimes based on their uses. 

Regulators should not expect to be able to rely solely on rigid, prescriptive rules, as if they could provide fixed ``guardrails'' that protect the public from the risks of generative AI and foundation models \cite{coglianese2025crum}. ``One size'' will decidedly not ``fit all'' when it comes to such multifunctional AI. Instead, regulators will need to consider more flexible strategies: performance standards, disclosure regulation, ex post liability, and management-based requirements. These more flexible approaches, when applied wisely and, as needed, in combination with one another, can give regulators a better grip on the use and problem heterogeneity of multifunctional AI. 

Experience with regulating heterogeneous processes and products in other realms suggests that successful AI governance will rest ultimately upon efforts to build human capital and organizational capacity—both within firms and across regulatory bodies. These efforts should aim to facilitate regulatory flexibility and an ability for rules and private risk management behavior to adapt to ever-changing environments. Of course, regulatory flexibility will itself become risky if not combined with regulatory vigilance and agility, so that appropriate action can be taken swiftly when needed to respond to new problems as they emerge. The road ahead demands the utmost in regulatory excellence—but just as with strategies for addressing the risks associated with other multifunctional technologies, meaningful, multi-faceted regulatory strategies do exist for responding to the risks of foundation models and generative AI. 

\clearpage
\bibliographystyle{ieee}
\bibliography{egbib}

\end{document}